\documentclass[aps,pra,twocolumn,amsmath,amssymb,footinbib,showpacs]{revtex4-1}

\newcommand{\Jnature}{Nature (London)}
\newcommand{\Jnatphys}{Nat. Phys.}

\newcommand{\Jprl}{Phys. Rev. Lett.}
\newcommand{\Jpr}{Phys. Rev.}
\newcommand{\Jpra}{Phys. Rev. A}
\newcommand{\Jprb}{Phys. Rev. B}

\newcommand{\Jrmp}{Rev. Mod. Phys.}

\newcommand{\Jepl}{Europhys. Lett.}
\newcommand{\Jnjp}{New J. Phys.}

\newcommand{\Jepjd}{Eur. Phys. J. D}
\newcommand{\Jepjst}{Eur. Phys. J. Special Topics}

\newcommand{\Jstatmech}{J. Stat. Mech.}

\newcommand{\Jadvphys}{Adv. Phys.}

\newcommand{\Jannphys}{Ann. Phys.}

\newcommand{\Jadvatmoloptphys}{Adv. At. Mol. Opt. Phys.}

\usepackage[english]{babel}
\usepackage{latexsym}
\usepackage{graphics}
\usepackage{subfigure}
\usepackage{epsfig}
\usepackage{color}
\usepackage{hyperref}
\hypersetup{
colorlinks=true,
citecolor=blue,
linkcolor=red,
urlcolor=black
}

\newcommand{\epsfboxmod}[1]{\epsfbox{#1}}

\newcommand{\infigbis}[2]{          \mbox{ \epsfxsize #1 \epsfboxmod{#2}}
                                      \vspace{-0.8cm}}

\newcommand{\ie}{{i.e.}}

\newcommand{\nc}{n}
\newcommand{\T}{T}
\newcommand{\J}{t}
\newcommand{\U}{U}

\newcommand{\ha}{\hat{a}}
\newcommand{\had}{\hat{a}^\dagger}

\newcommand{\re}{\rho_E}
\newcommand{\gp}{g^+_j}
\newcommand{\gm}{g^-_j}
\newcommand{\Vrond}{\mathcal{V}^{E}_j}

\begin{document}

\title{Localization transition in weakly interacting Bose superfluids in one-dimensional quasiperdiodic lattices}

\author{Samuel Lellouch}
\author{Laurent Sanchez-Palencia}
\affiliation{
  Laboratoire Charles Fabry,
  Institut d'Optique, CNRS, Univ Paris Sud 11,
  2 avenue Augustin Fresnel,
  F-91127 Palaiseau cedex, France
}

\date{\today}

\begin{abstract}
We study the localization of collective pair excitations
in weakly-interacting Bose superfluids in one-dimensional quasiperiodic lattices.
The localization diagram is first determined numerically.
For intermediate interaction and quasiperiodic amplitude
we find a sharp localization transition,
with extended low-energy states and localized high-energy states.
We then develop an analytical treatment, which allows us to quantitatively map the localization transition into that of an effective multiharmonic quasiperiodic system.
\end{abstract}


\maketitle

Quasiperdiodic systems, which are formed of a small number of incommensurate sinusoidal components,
constitute an appealing intermediate between disordered and periodic systems.
Such structures are basic models for a wide variety of physical systems.
They appear naturally in the growth of certain crystals~\cite{shechtman1984}
or as a result of charge-density waves~\cite{wilson1975}.
They also describe two-dimensional lattice electrons in perpendicular magnetic fields~\cite{peierls1933,*harper1955,*hofstadter1976}.
Moreover, they can be created on purpose
in solid crystals~\cite{merlin1985},
photonic crystals~\cite{lahini2009},
and ultracold-atom optical lattices~\cite{guidoni1997,*roth2003,*lsp2005,fallani2007,*gadway2011,*tanzi2013,*derrico2014,fallani2008,*modugno2009,*lsp2010}.
In quasiperiodic systems, the lack of translation invariance can induce localization
of linear waves, similarly as the phenomenon of Anderson localization in disordered systems~\cite{anderson1958}.
In quasiperiodic systems, however, the quasi-repetition of finite patterns radically changes
the localization picture.
For instance, in a one-dimensional disordered system, any quantum particle is localized with
an energy-dependent localization length~\cite{abrahams1979,lifshits1988}.
In contrast, for a quasiperiodic system made of a single incommensurate sinusoidal modulation
of a main periodic lattice, there is a localization transition for some critical strength of the quasiperiodic component beyond which the states are localized with a localization length that is independent of the energy~\cite{aubry1980,*sokoloff1981,*aulbach2004}.

The extension of the concept of localization to interacting quantum systems is attracting a considerable
attention as regards
phase diagrams~\cite{giamarchi1987,*giamarchi1988},
many-body localization transitions~\cite{basko2006,*oganesyan2007,*aleiner2010,*pal2010},
and localization of collective excitations~\cite{gurarie2002,*gurarie2003,bilas2006,lugan2007b,*lugan2011}.
These issues have been first investigated for purely disordered systems
and extensions to quasiperiodic systems are just starting.
So far, most studies focused on the phase diagram of one-dimensional bosons in quasiperiodic lattices
at zero temperature~\cite{roscilde2008,roux2008,roux2013},
finite temperature~\cite{michal2014},
and infinite temperature~\cite{iyer2013}.
Conversely, the localization of collective excitations remains largely open.
This issue is particularly important because the transport of collective excitations governs many dynamical effects in correlated quantum systems~\cite{polkovnikov2011},
for instance the propagation of correlations in recently-developed quench experiments~\cite{cheneau2012,*trotzky2012,*langen2013}.

Here we study the localization of collective pair excitations
in weakly interacting Bose superfluids subjected to a one-dimensional quasiperiodic lattice.
We first determine the localization diagram numerically and show that,
for intermediate interaction and quasiperiodic amplitude,
there is a sharp localization transition.
This nontrivial transition separates bands of states that are extended at low energy and localized at high energy.
We then develop an analytical treatment, which allows us to reproduce the numerical results
accurately
and to quantitatively map the localization transition into that of an effective multiharmonic quasiperiodic system.
Finally, we discuss experimental observability and possible extension of our results.

The starting point of our study is the Aubry-Andr\'e-Hubbard Hamiltonian,
 \begin{equation}
\hat{H}=-\sum_{j,l}\T_{j,l}\had_j\ha_l + \sum_j V_j\had_j\ha_j + \dfrac{\U}{2}\sum_j \had_j\had_j\ha_j\ha_j,
 \label{eq:Hamiltonian}
 \end{equation}
which governs the low-energy physics of interacting bosons in one-dimensional (1D) quasiperiodic lattices.
In Eq.~(\ref{eq:Hamiltonian}), $\ha_j$ and $\had_j$ are the bosonic annihilation and creation operators at the lattice site $j$.
The first term represents quantum tunneling with
the hopping matrix $\hat{T}$,
which includes nearest-neighbor tunneling, $\T_{j,j\pm 1}=\J$ and $\T_{j,l}=0$ for $|j-l|>1$,
as well as the homogeneous on-site term $\T_{j,j}=-2\J$, for convenience.
Within this convention, the free-particle spectrum, $\varepsilon_k=4\J\sin^2(k/2)$,
is centered on $\varepsilon=2\J$ with the band edges $\varepsilon=0$ and $\varepsilon=4\J$.
The second term represents the on-site quasiperiodic potential modulation,
$V_j = \Delta \cos(2\pi r j + \varphi)$,
where
$\varphi$ is a phase,
$\Delta$ is the quasiperiodic amplitude, and
$r$ is an irrational number.
The third term represents on-site repulsive interactions with the interaction energy $\U>0$.

In the weakly interacting regime
with high occupation number per lattice site ($n\gg U/\J$, with $n$ the mean density),
we can rely on mean field theory~\cite{vanoosten2001}.
A similar approach has been presented elsewhere for disordered systems in continuous~\cite{lsp2006,lugan2007b,*lugan2011,gaul2011b} or lattice~\cite{gaul2013} spaces,
and we just outline it here.
The density background $n_j$ is first determined by minimizing the classical energy functional,
obtained by replacing the operator $\ha_j$ by the real-valued field $\phi_j \equiv \sqrt{n_j}$ in Eq.~(\ref{eq:Hamiltonian}).
It yields the Gross-Pitaevskii equation (GPE),
\begin{equation}
\mu\phi_j=-\hat{T}\phi_j+V_j\phi_j+\U\phi_j^{3},
\label{eq:GPE}
\end{equation}
where the term $\hat{T}\phi_j$ is a shortcut for the hopping matrix contribution
$\J(\phi_{j+1}-2\phi_{j}+\phi_{j-1})$
and $\mu$ is the chemical potential.
The collective pair excitations of the Bose superfluid, which are represented by two fields $u_j$ and $v_j$,
are then found by expanding Hamiltonian~(\ref{eq:Hamiltonian}) up to second order in the Bogoliubov operator $\delta \hat{n}_j /2\sqrt{n_j} + i\sqrt{n_j} \delta \hat{\theta}_j$,
where $\delta \hat{n}_j$ and $\delta \hat{\theta}_j$ are the density and phase fluctuation operators, and diagonalizing the resulting quadratic Hamiltonian.
The excitation energy $E$ and wave functions $u_j$ and $v_j$ are the solutions
of the Bogoliubov-de Gennes equations (BdGEs)
\begin{eqnarray}
\left[ \begin{matrix}
-\hat{T} \! + \! V_j \! - \! \mu \! + \! 2\U n_j
& \U n_j \\
-\U n_j
& \hat{T} \! - \! V_j \! + \! \mu \! - \! 2\U n_j  \end{matrix} \right]
\! \left[ \begin{matrix} u_j  \\ v_j  \end{matrix} \right]
\! = \!
 E \! \left[ \begin{matrix} u_j  \\ v_j \end{matrix} \right]\!.~~~~~
\label{eq:BdGE}
\end{eqnarray}
Equations~(\ref{eq:GPE}) and (\ref{eq:BdGE}) form the complete set to determine
the elementary excitations of the Bose fluid in the quasiperiodic lattice.

We first solve Eqs.~(\ref{eq:GPE}) and (\ref{eq:BdGE}) numerically in the 1D quasiperiodic lattice.
The number of lattice sites is chosen to be a Fibonacci number $F_p$ and $r$ is taken as the ratio $F_{p-1}/F_p$. It allows us to use periodic boundary conditions and a good approximation
of an incommensurate ratio $(\sqrt{5}-1)/2$~\cite{aulbach2004}.
In practice, we use $F_p=610$, which yields  $r=(\sqrt{5}-1)/2 \pm 0.000002$~\footnote{We have checked that the results do not depend on the system size by using other numbers of lattice sites, for instance $F_p=233$ and $F_p=987$.}.
The density background is computed by solving the GPE using imaginary-time propagation with a Crank-Nicolson scheme~\cite{NumericalRecipes}.
The good numerical convergence of the imaginary-time propagation of the GPE is a delicate point for the subsequent determination of the collective excitations using Eq.~(\ref{eq:BdGE}).
The convergence criterion applies to the effective, imaginary-time-dependent chemical potential
$\mu(\tau) \equiv -\frac{\hbar}{2}\frac{d}{d \tau}\ln \left(\sum_j n_j\right)$. We have checked that the density profile is unaffected when the precision threshold varies from $10^{-8}$ to $10^{-15}$. The same holds when the imaginary time step $\Delta\tau$ used in the propagation varies from $0.01/\J$ to $0.5/\J$.
Moreover, the density profile precisely agrees with the perturbative expansion of the GPE solution implemented up to order 50~\cite{note:supplement}.
All together, the precision on the density profile $n_j$ is of the order of $10^{-8}$ for all results presented here. The excitations are then computed by exact diagonalization of the matrix in Eq.~(\ref{eq:BdGE})
using the Lanczos algorithm for sparse non-Hermitian eigenproblems~\cite{NumericalRecipes}.

\begin{figure}[t!]
\begin{center}
 \infigbis{27em}{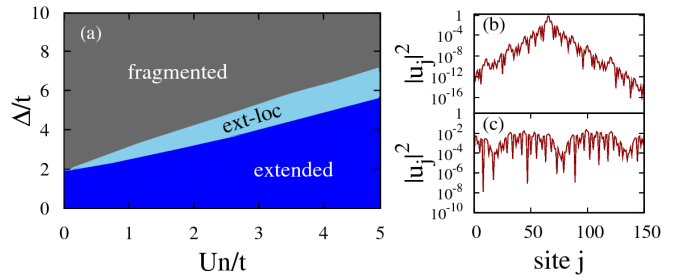}
\end{center}
\vspace{-0.2cm}
\caption{(Color online) Numerical results.
(a)~Localization diagram as a function of the interaction strength and the quasiperiodic amplitude.
It displays three regimes:
(i)~`extended regime' where the density background is connected and all excitations are extended;
(ii)~`fragmented regime' where the density background is fragmented; and
(iii)~`extended-localized regime' where the density background is connected and the excitation spectrum shows a delocalization-localization transition with
exponentially localized high-energy states
and extended low-energy states.
(b)-(c)~Typical excitation wave function $u$ in the  localized (b) and extended (c) regimes,
plotted in semilogarithmic scale and for the $150$ first lattice sites (similar plots are found for the $v$ wave functions).
The two panels correspond to two excitations with consecutive energies
above (b) and below (c) the mobility edge for $U\nc/\J=1.75$ and $\Delta/\J=3.3$.
\label{fig:f1}
}.
\end{figure}

The numerical results are summarized on the diagram in Fig.~\ref{fig:f1}(a).
It displays three different regimes.
For weak quasiperiodic amplitude $\Delta$ and strong interaction $U$,
the density background is fully connected and all excitations are extended (`extended regime').
For a given interaction strength $U$ and tunneling $\J$, the density modulations increase
with the quasiperiodic amplitude $\Delta$.
Above a critical value of $\Delta_\textrm{\tiny c}$, 
the density profile gets fragmented (`fragmented regime'), which yields the upper boundary on the diagram.
The fragmentation condition is chosen to be the minimal value of $\Delta$ such that at least one lattice site
has a density lower than $0.01$ atom per site.
We have checked that varying this arbitrary threshold down to $0.001$ yields insignificant changes
of the fragmentation boundary. 
Moreover the latter is in good agreement with the experimental observation of Ref.~\cite{tanzi2013}.
In the fragmented regime, the density profile is cut in disconnected pieces.
It corresponds to trivial localization, a case that we disregard in the following.
Notice that in the limit $U \rightarrow 0$, we recover the critical value
$\Delta_\textrm{\tiny c}=2\J$, which is the localization transition of the noninteracting Aubry-Andr\'e model.
The most interesting regime appears for intermediate quasiperiodic amplitude (`ext-loc regime').
In this regime, although the density background is fully connected,
we find a localization transition of the collective excitations.
Remarkably enough, they are the high-energy excitations that are exponentially localized over a few lattice sites [see Fig.~\ref{fig:f1}(b)] while the low-energy excitations are extended over the whole system [see Fig.~\ref{fig:f1}(c)].
This transition is sharp as exemplified in Figs.~\ref{fig:f1}(b) and (c),
which correspond to two excitations of consecutive energies for $U n=1.75\J$ and $\Delta=3.3\J$.

\begin{figure}[t!]
\begin{center}
 \infigbis{24em}{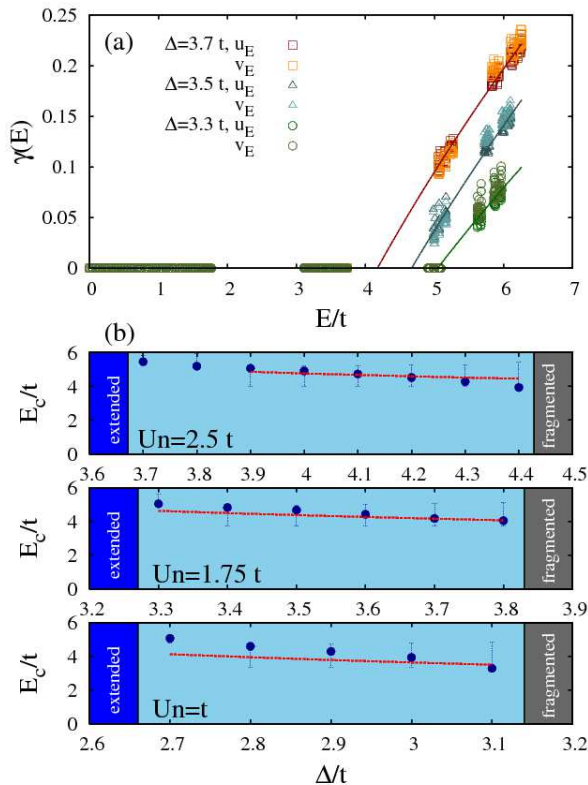}
\end{center}
\caption{(Color online) (a) Lyapunov exponents of the Bogoliubov wave functions $u$ and $v$,
for $Un/\J=1.75$ and $\Delta/\J=3.3, 3.5, 3.7$.
The excitation spectrum is banded and displays a sharp localization transition separating extended ($\gamma=0$) and localized ($\gamma>0$) states.
(b) Mobility edge as a function of the quasiperiodic amplitude $\Delta/\J$ as extracted from 
power-law fits to the numerical $\gamma(E)$ curves [solid lines on panel~(a)]. 
Error-like bars correspond to the edges of the minigap containing the mobility edge.
The dotted, red line shows the analytical prediction of the locator theory
applied to the effective model~(\ref{eq:Schrodeff}) with the potential~(\ref{eq:Vrond2}).
\label{fig:f2}
}
\end{figure}

In order to characterize the localization transition,
we compute two Lyapunov exponents for the excitations, which
correspond to the two Bogoliubov wave functions,
$\gamma_{u}(E) \equiv -\lim_{j\rightarrow \infty} \ln |u_j|/j$
and
$\gamma_{v}(E) \equiv -\lim_{j\rightarrow \infty} \ln |v_j| /j$.
They are extracted from fits in the tails of the logarithm of the wave functions $u$ and $v$.
Figure~\ref{fig:f2}(a) displays those Lyapunov exponents versus the excitation energy $E$,
for fixed interaction and disorder strengths.
The Lyapunov exponents $\gamma_{u}$ and $\gamma_{v}$ are indistinguishable
and hereafter we omit the wave function index $u$ or $v$.
In the `ext-loc' regime the Lyapunov exponent curves clearly show the transition,
separating extended ($\gamma=0$) and localized ($\gamma>0$) states~\footnote{The same trend can be found on the excitation participation ratio~\cite{roux2013}. We have also computed it and, comparing the behaviors of the participation ratio and the Lyapunov exponent, we found that the latter permits a better quantitative estimation of the localization transition.}.
The excitation spectrum splits in several bands separated by minigaps, a general feature in quasiperiodic systems~\cite{sokoloff1980,sokoloff1981,soukoulis1982,johansson1991,aulbach2004,biddle2011}.
The transition generally lies in one of the minigaps. 
To determine the mobility edge $E_\textrm{c}$,
we thus rely on fits of the $\gamma(E)$ curves with several fitting functionals
[linear, $\gamma(E) \sim E-E_\textrm{c}$;
power-law, $\gamma(E) \sim E^\alpha-E_\textrm{c}^\alpha$;
and
logarithmic, $\gamma(E) \sim \ln(E/E_\textrm{c})$].
The result is found to be almost independent of the fitting functional
and thus provides a reliable estimate of the mobility edge.
Figure~\ref{fig:f2}(b) shows the mobility edge versus the quasiperiodic amplitude
for various interaction strengths.
The errorlike bars represent the edges of the minigap containing the mobility edge.
The uncertainty on the fitted mobility edge is smaller than these bars.

In order to interpret those results, we now turn to an analytical treatment of the localization problem.
The main difficulty relies on the fact that localization in quasiperiodic systems
occurs for strong quasiperiodic amplitude $\Delta$~\cite{aubry1980}.
For this reason, the lowest-order perturbation theory, 
which proved successful for 1D disordered systems~\cite{bilas2006,lugan2007b,*lugan2011},
fails here~\footnote{We have found that lowest-order perturbation theory as used
for 1D disordered, interacting Bose gases in Refs.~\cite{bilas2006,lugan2007b,*lugan2011}
is both quantitatively and qualitatively incorrect. In particular, it
predicts a diagram that is inconsistent with that of Fig.~\ref{fig:f1}
and it is not able to predict the localization transition.}.
To overcome this issue, we develop an approach based on a generic expansion in
harmonics of the quasiperiodic potential.
The structure of the GPE~(\ref{eq:GPE}) shows that the field $\phi_j$ takes the form of a series of harmonics of the quasiperiodic potential. The density field $n_j$ thus reads $n_j=(\mu-\tilde{V}_j)/U$, where
\begin{equation}
\tilde{V}_j=\sum_{p \geq 1} A_p \cos [p(2\pi r j + \varphi)]
\label{eq:VtildeSeries}
\end{equation}
is a multiharmonic quasiperiodic field,
the coefficients of which can be computed iteratively~\cite{note:supplement}.
Using the energy-dependent linear transform~\cite{lugan2007b,*lugan2011}
\begin{equation}
g^{\pm}_{j}=\pm \re^{\pm 1/2}(u_j-v_j)+\re^{\mp 1/2}(u_j+v_j)
\label{eq:gpm}
\end{equation}
where $\re = \sqrt{1+(\mu/E)^2} + \mu/E$, the BdGEs~(\ref{eq:BdGE}) exactly rewrite
\begin{eqnarray}
-(\re^{-1} E + \hat{T}) \gp  + \! \Bigg[V_j \! - \! \dfrac{3+\re^2}{1+\re^2}\tilde{V}_j  \Bigg]\gp
& = &
\dfrac{2\re\tilde{V}_j}{1+\re^2}  \gm
\label{eq:BdGEgp} \\
(\re E - \hat{T})\gm  + \! \Bigg[V_j \! - \! \dfrac{1+3\re^2}{1+\re^2}\tilde{V}_j \Bigg]\gm 
& = &
\dfrac{2\re \tilde{V}_j}{1+\re^2}  \gp.~~~~
\label{eq:BdGEgm}
\end{eqnarray}
The solution of these equations is significantly simplified by noticing that
the lattice-space Green function of the operator $-\hat{T}+\re E$ is of width $\sqrt{\J/\re E}$ and amplitude $1/(\re E + 2\J)\sqrt{1-[2\J/(\re E + 2\J)]^2}$.
Hence, for $\re E \gg \J$, this operator can be replaced by the local operator
$\re E + 2\J$ in Eq.~(\ref{eq:BdGEgm}).
It is then straightforward to write the expression of $\gm$ as a function of $\gp$ and of the potentials $V_j$ and $\tilde{V}_j$.
Inserting this expression into Eq.~(\ref{eq:BdGEgp}) we find a closed equation for $\gp$,
 \begin{equation}
-\hat{T} \gp + \Vrond \gp = E\re^{-1} \gp,
 \label{eq:Schrodeff}
 \end{equation}
with the effective potential
\begin{equation}
\Vrond \simeq V_j - \dfrac{3+\re^2}{1+\re^2}\tilde{V}_j
        -\dfrac{\left(\dfrac{2\re}{1+\re^2}\right)^2 \tilde{V}_j^2}{\re E+2\J+V_j-\dfrac{1+3\re^2}{1+\re^2}\tilde{V}_j}.
\label{eq:Vrond}
\end{equation}
Using exact diagonalization of Eq.~(\ref{eq:Schrodeff}) with the potential~(\ref{eq:Vrond}) around energy $E$,
we have checked that
the Lyapunov exponents and the localization transition given by our effective model coincide
with those found using direct diagonalization of the BdGEs~(\ref{eq:BdGE}).
It validates the effective model~(\ref{eq:Schrodeff})-(\ref{eq:Vrond})
and the approximation $-\hat{T}+\re E \simeq 2\J + \re E$ used above.

In this model, the quantity $\Vrond$ is a multiharmonic periodic potential of spacing $1/r$ incommensurate with that of the main lattice, which is unity.
Such systems are known to exhibit in general an energy-dependent mobility edge with low-energy extended states and high-energy localized states~\cite{sokoloff1980,soukoulis1982,johansson1991,biddle2011}. This holds except in the particular case of self-dual models, among which the Aubry-Andr\'e model is a celebrated example~\cite{aubry1980}.
Self-duality requires a specific relation between the amplitudes of the $p$th harmonics and of the tunneling rate to the $p$th neighbors. The latter does not apply in our case since tunneling is strictly restricted to the first neighbors. It qualitatively explains the localization transition of the collective excitations reported here.

Localization properties in quasiperiodic systems can be further inferred from locator perturbation theory~\cite{sokoloff1980}. Here the localization criterion roughly corresponds to the convergence of the self-energy in the thermodynamic limit, which reads $D(E)>1$ where $D(E)$ is the so-called localization function.
In the case of Eq.~(\ref{eq:Schrodeff}), it reads
\begin{equation}
D(E) = \exp\left( r \int_0^{1/r} \!\!\! dx\, \ln \left| \frac{E\re^{-1}-2\J-\mathcal{V}^{E}(x)}{\J}\right| \right).
\label{eq:D(E)}
\end{equation}
Equation~(\ref{eq:D(E)}) can, in principle, be applied to the full effective potential $\mathcal{V}^{E}(x)$.
To obtain analytical results, it is, however, worth truncating the infinite series of harmonics in $\mathcal{V}^E$.
Keeping only one harmonic is not sufficient to capture the physics even qualitatively,
since it would unphysically restore duality and change the universality class of the localization transition.
On the other hand, beyond two, the number of harmonics does not change the universality class.
We may thus restrict ourselves to the two lowest-order harmonics,
which are generated in first instance in second-order perturbation theory.
As we shall see, this order of expansion turns out to be sufficient for a significant part of the localization diagram of Fig.~\ref{fig:f1}.
It yields the effective two-harmonic potential
\begin{equation}
\Vrond \simeq \Delta_E^{(0)} + \Delta_E^{(1)}\cos(2\pi r j+\varphi) + \Delta_E^{(2)}\cos[2(2\pi r j+\varphi)]
\label{eq:Vrond2}
\end{equation}
with the amplitudes
\begin{eqnarray}
\Delta_E^{(0)} & = & \dfrac{3 \! + \! \re^2}{1 \! + \! \re^2} \dfrac{\Delta^2}{4Un}(f_r \! - \! f_r^2)-\dfrac{2\re^2}{(1 \! + \! \re^2)^2}\dfrac{\Delta^2 f_r^2}{\re E \! + \! 2\J}
\label{eq:D0} \\
\Delta_E^{(1)} & = & \Delta \Bigg[1- \dfrac{3+\re^2}{1+\re^2}f_r \Bigg] 
\label{eq:D1} \\
\Delta_E^{(2)} & = & \dfrac{3+\re^2}{1+\re^2} \dfrac{\Delta^2}{4Un}
\Bigg[\frac{f_r^2}{2} + \left(f_r-\frac{3}{2}f_r^2\right)f_{2r}\Bigg] \nonumber\\
& & -\dfrac{2\re^2}{(1+\re^2)^2}\dfrac{\Delta^2 f_r^2}{\re E+2\J}.
\label{eq:D2}
\end{eqnarray}
and $f_r=\dfrac{1}{1+2\J\sin^2(\pi r)/Un}$~\cite{note:supplement}.
The results of the locator perturbation theory applied to the two-harmonic potential~(\ref{eq:Vrond2})
is shown in Fig~\ref{fig:f2}(b). 
It predicts the correct localization transition with collective excitations that are extended at low energy and localized at high energy, and a mobility edge that is in very good agreement with the full numerical result.
Perturbation theory beyond second order generates high-order harmonics in the effective potential
and renormalizes the amplitudes $\Delta^{(p)}$.
For $\U n \lesssim 3 \J$, we find that they induce negligible effects and do not significantly affect the prediction for the mobility edge.
For higher values of $\U$, however, second-order perturbation theory is not sufficient to accurately estimate the density background, and higher-order terms should be included.

In summary, we have shown that the collective excitations of lattice Bose superfluids subjected
to a single-harmonic quasiperiodic potential undergo a nontrivial localization transition
with extended low-energy states and localized high-energy states.
Therefore the interactions change the universality class of the localization transition,
in striking contrast with the purely disordered case~\cite{bilas2006,lugan2007b,lugan2011}.
In the quasiperiodic case the transition can be understood as the result of the scattering of the excitations
from the potential and the density background, which contains an infinite series of harmonics of the potential.
It could be observed in ultracold-atom experiments,
using for instance spectroscopy techniques, which give direct access to the excitations~\cite{stenger1999,*steinhauer2002,*richard2003},
or in quench experiments, which generate collective excitations that govern the propagation of experimentally-observable correlations~\cite{calabrese2006,*lauchli2008,*carleo2014,cheneau2012,*trotzky2012,*langen2013}.
It could also be observed in photonic crystals, which can combine quasiperdiodic structures~\cite{lahini2009} and photon nonlinearities~\cite{lahini2008}. Finally, it would be interesting to study the counterpart of the localization transition discussed here in Fermi superconductors, which may directly apply to electronic quasicrystals.

We are grateful to Giovanni Modugno and Massimo Inguscio for insightful discussions.
This research was supported by
the European Research Council (FP7/2007-2013 Grant Agreement No.\ 256294),
the Minist\`ere de l'Enseignement Sup\'erieur et de la Recherche,
and
the Institut Francilien de Recherche sur les Atomes Froids (IFRAF).
We acknowledge the use of the computing facility cluster GMPCS of the LUMAT federation (FR LUMAT 2764)
and HPC resources from GENCI-IDRIS (Grants 2013057143 and 2014057143).

%

\renewcommand{\theequation}{S\arabic{equation}}
\setcounter{equation}{0}
\renewcommand{\thefigure}{S\arabic{figure}}
\setcounter{figure}{0}
\onecolumngrid  
    
\pagebreak

\newpage

{\center \bf \large Supplemental Material\vspace*{1.cm}\\}

This supplemental material aims at providing some details about the harmonic structure of the density profile and the effective potential $\Vrond$, as well as the series expansion used to determine them.

\subsubsection{Series expansion of the density background}

To determine the density background $n_j \equiv \phi_j^2$, we solve the Gross-Pitaevskii equation (GPE) together with the normalization condition $n=\frac{1}{L}\Sigma_j n_j$ where $L$ is the number of sites and $n$ is the averaged density. To do so, we perform a series expansion in powers of the quasiperiodic potential $V_j$.
In the absence of an external potential, we have $\phi_j=\sqrt{\nc}$ and the chemical potential $\mu=U\nc$.
In the presence of an external potential, we then write
\begin{eqnarray}
\phi_j & = & \sqrt{\nc} \left( \phi^{(0)}_j + \phi^{(1)}_j + \phi^{(2)}_j ... \right)
\label{eq:ExpPhi} \\
\mu & = & U\nc \left( \mu^{(0)}+\mu^{(1)}+\mu^{(2)}+... \right)
\label{eq:ExpMu}\\
n_j & = & \nc \left( n^{(0)}_j + n^{(1)}_j + n^{(2)}_j ... \right)
\label{eq:Expn} 
\end{eqnarray}
where the superscripts denote increasing orders in the quasiperiodic amplitude $\Delta$,
and $\phi^{(0)}_j=1$, $n^{(0)}_j=1$, $\mu^{(0)}=1$.
Notice that the chemical potential has to be expanded also to fulfill the normalization condition.
Inserting the expansions~(\ref{eq:ExpPhi}) and (\ref{eq:ExpMu}) in the GPE [Eq.~(\ref{eq:GPE}) of the paper] and the expansion~(\ref{eq:Expn}) in the normalization condition, we get
\begin{equation}
U\nc\left(\mu^{(0)}+\mu^{(1)}+...\right)(\phi^{(0)}_j + \phi^{(1)}_j+...) = - \hat{T}\left(\phi^{(0)}_j + \phi^{(1)}_j + ...\right) +V_j\left(\phi^{(0)}_j + \phi^{(1)}_j + ...\right) +U\nc\left(\phi^{(0)}_j + \phi^{(1)}_j +  ...\right)^{3}
 \label{eq:ExpGPE}
\end{equation}
and
\begin{equation}
\frac{1}{L}\sum_j \left(\phi^{(0)}_j + \phi^{(1)}_j + ...\right)^2 = 1.
\label{eq:ExpNor}
\end{equation}
Then, collecting all the terms of same order $p$ in the quasiperiodic amplitude yields
\begin{equation}
\left(1-\frac{1}{2U\nc}\hat{T}\right)\phi_j^{(p)}
=
-\dfrac{V_j}{2U\nc}\phi_j^{(p-1)}
-\dfrac{1}{2}\sum_{k,\ell,m=p,0\leq k,\ell,m \leq p-1} \phi_j^{(k)}\phi_j^{(\ell)}\phi_j^{(m)} + \dfrac{1}{2}\sum_{1\leq k \leq p-1}\mu^{(k)}\phi_j^{(p-k)} + \dfrac{\mu^{(p)}}{2}
 \label{eq:GPEordern}
\end{equation}
and
\begin{equation}
\sum_j \left(2\phi^{(0)}_j\phi^{(p)}_j + \sum_{1\leq k \leq p-1} \phi^{(k)}_j\phi^{(p-k)}_j\right) = 0.
\label{eq:Norordern}
\end{equation}
Equations~(\ref{eq:GPEordern}) and (\ref{eq:Norordern}) can then be used to compute all $\phi^{(p)}_j$ and $\mu^{(p)}$ at any order $p$ iteratively.
The iteration process works as follows.
Given all $\phi^{(k)}_j$ and $\mu^{(k)}$ at orders $k < p$, we calculate $\phi^{(p)}_j$ as a function of $\mu^{(p)}$ from Eq.~(\ref{eq:GPEordern}) by inverting the operator $1-\hat{T}/{2U\nc}$.
The quantity $\mu^{(p)}$ is then found by inserting this expression for $\phi^{(p)}_j$ into Eq.~(\ref{eq:Norordern}).
Having determined $\phi^{(p)}_j$, we then find the density field using Eq.~(\ref{eq:Expn}),
whose expansion in powers of the quasiperiodic amplitude writes
\begin{equation}
n^{(p)}_j= \sum_{0\leq k,\ell \leq p,k+\ell=p}\phi^{(k)}_j\phi^{(\ell)}_j.
\label{eq:nn}
\end{equation}
 
\bigskip
This procedure is completely general and can be applied to any external potential $V_j$. 
In the case of the quasiperiodic potential $V_j=\Delta \cos(2\pi r j + \varphi)$, the above iterative process is fully algebraic because the operator $1-\hat{T}/{2U\nc}$ in Eq.~(\ref{eq:GPEordern}) can be analytically inverted at any order (see below). 
We have implemented this expansion up to order 50 and found excellent agreement with
the direct numerical solution of the GPE~(\ref{eq:GPE}). It provides a cross-check of
the precision of the numerical solution and of the convergence of the present analytical expansion.

\subsubsection{Analytical expansion in the case of a quasiperiodic potential}
We now give some explicit formulas for the lowest order terms and discuss the harmonic structure of the density background $n_j$.
As in the paper, we generically write the density field $n_j$ 
\begin{equation}
n_j=(\mu-\tilde{V}_j)/U,
\label{eq:Vtildedef}
\end{equation}
where the field $\tilde{V}$ includes terms of all orders.

\bigskip
At first order, Eq.~(\ref{eq:GPEordern}) reduces to $-\hat{T}\phi_j^{(1)}+2U\nc\phi_j^{(1)}=-V_j+U\nc\mu^{(1)}$.
It is straightforward to solve it in Fourier space where the operator $\hat{T}$ is diagonal. It yields
$\phi^{(1)}_k=-\dfrac{V_k-Un\mu^{(1)}\delta_{k,0}}{\varepsilon^0_k+2U\nc}$
where $\varepsilon^0_k = 4\J\sin^2(k/2)$.
Inserting this expression into Eq.~(\ref{eq:Norordern}), we find $\mu^{(1)}=V_{k=0}/U\nc=0$ and $\phi^{(1)}_k=-\dfrac{V_k}{\varepsilon^0_k+2U\nc}$. Remarkably, since the quasiperiodic potential contains only one spatial frequency, $V_k=\Delta (e^{i\varphi}\delta_{k,+2\pi r}+e^{-i\varphi}\delta_{k,-2\pi r})/2$, one can immediately get back to real space
and write 
\begin{equation}
\phi^{(1)}_j=-\dfrac{\Delta}{2U\nc}f_r \cos(2\pi r j+\varphi),
\label{eq:psi1}
\end{equation}
where $f_r=\dfrac{1}{1+\varepsilon^0_{2\pi r}/2U\nc}$.
Hence, to lowest order, the density profile is quasiperiodic field.
It follows the modulations of the quasiperiodic potential with a reduced amplitude since $f_r<1$.
The factor $f_r$ is a remainder of the nonlocal operator $1-\hat{T}/2Un$ in the l.h.s.\ of Eq.~(\ref{eq:GPEordern}), which reduces to an algebraic operation in the case of a quasiperiodic potential.
Then, Eqs.~(\ref{eq:nn}) and (\ref{eq:Vtildedef}) yield the first order term of the field $\tilde{V}_j$. It reads $\tilde{V}^{(1)}_j=-2Un\phi^{(1)}_j$ where $\phi^{(1)}_j$ is given by Eq.~(\ref{eq:psi1}), \ie\
\begin{equation}
\tilde{V}^{(1)}_j = \Delta f_r \cos(2\pi r j+\varphi).
\label{eq:VtildeO1}
\end{equation}

\bigskip
The next orders are found following the same process, which remains algebraic to any order
in the case of the quasiperiodic potential. To second order, it yields the term
\begin{equation}
\phi^{(2)}_j=\left(\dfrac{\Delta}{2U\nc}\right)^{2}\Bigg[-\dfrac{f_r^2}{4} + \left(f_r-\frac{3}{2}f_r^2\right)f_{2r}\dfrac{\cos[2(2\pi r j+\varphi)]}{2}\Bigg],
\label{eq:psi2}
\end{equation}
and a negative shift on the chemical potential,
\begin{equation}
\mu^{(2)}=-\Bigg(\dfrac{\Delta}{2U\nc}\Bigg)^{2}(f_r-f_r^2).
\label{eq:mu2}
\end{equation}
The field $\tilde{V}_j$ is then given at second order by $\tilde{V}^{(2)}_j=Un[\mu^{(2)}-n^{(2)}_j]=Un[\mu^{(2)}-(2\phi^{(2)}_j+\phi^{(1)2}_j)]$ where $\phi^{(1)}$, $\phi^{(2)}$ and  $\mu^{(2)}$ are given by Eqs.~(\ref{eq:psi1}), (\ref{eq:psi2}), and (\ref{eq:mu2}), \ie\
\begin{equation}
\tilde{V}^{(2)}_j = - \frac{\Delta^2}{4Un} \left\{f_r - f_r^2 + \left[ \frac{f_r^2}{2} + \left(f_r-\frac{3}{2}f_r^2 \right) f_{2r} \right] \cos[2(2\pi r j+\varphi)] \right\}.
\label{eq:VtildeO2}
\end{equation}
Hence, the second-order terms $\phi^{(2)}_j$ and $\tilde{V}^{(2)}_j$ contain a constant term and the second harmonics of the quasiperiodic potential. Those terms are generated by the nonlinear term of the GPE:
Since the first order term contains only the first harmonics,
$\phi^{(1)}_j\propto\cos(2\pi r j+\varphi)$,
the product terms $\phi_j^{(1)}\phi_j^{(1)}\phi_j^{(0)}$ appearing in Eq.~(\ref{eq:GPEordern}) contain the zeroth and second harmonics.

\bigskip
More generally, it is straightforward to show recursively that the terms of order $p$, $\phi_j^{(p)}$ and $\tilde{V}^{(p)}_j$, contain the $p$th harmonics of the quasiperiodic potential, $\cos[p(2\pi r j+\varphi)]$, as well as all lower harmonics of same parity.
In particular, a constant term in $\phi_j$ and a correction to the chemical potential $\mu$ appear only at even orders.
Hence, the field $\tilde{V}_j$ takes the multiharmonic quasiperiodic form
\begin{equation}
\tilde{V}_j=\sum_p A_p \cos [p(2\pi r j + \varphi)],
\label{eq:psiSeries2}
\end{equation}
where the amplitude $A_p$ of the $p$th harmonics is a power series of order $p$,
$A_p\sim \alpha_p(\Delta/2U\nc)^p + \alpha_{p+2}(\Delta/2U\nc)^{p+2}+....$

\subsubsection{Expansion of the effective potential $\Vrond$}
The effective potential
\begin{equation}
\Vrond \simeq V_j - \dfrac{3+\re^2}{1+\re^2}\tilde{V}_j
        -\dfrac{\left(\dfrac{2\re}{1+\re^2}\right)^2 \tilde{V}_j^2}{\re E+2\J+V_j-\dfrac{1+3\re^2}{1+\re^2}\tilde{V}_j},
\label{eq:Vrondappen}
\end{equation}
which appears in the Eq.~(\ref{eq:Schrodeff}) of the paper,
can as well be expanded in powers of the quasiperiodic amplitude
by expanding the denominator and using the previously obtained expansions for $\tilde{V}_j$.
Up to second order, we get
\begin{equation}
\Vrond \simeq V_j - \dfrac{3+\re^2}{1+\re^2}\tilde{V}^{(1)}_j - \dfrac{3+\re^2}{1+\re^2}\tilde{V}^{(2)}_j
        -\dfrac{\left(\dfrac{2\re}{1+\re^2}\right)^2}{\re E+2\J} \left(\tilde{V}^{(1)}_j\right)^2,
\label{eq:Vrondorder2}
\end{equation}
where $\tilde{V}^{(1)}$ and $\tilde{V}^{(2)}$ are given in Eqs.~(\ref{eq:VtildeO1}) and (\ref{eq:VtildeO2}).
It yields the two-harmonic effective potential
\begin{equation}
\Vrond \simeq \Delta_E^{(0)} + \Delta_E^{(1)}\cos(2\pi r j+\varphi) + \Delta_E^{(2)}\cos[2(2\pi r j+\varphi)]
\label{eq:Vrond2appen}
\end{equation}
with the amplitudes
\begin{eqnarray}
\Delta_E^{(1)} & = & \Delta \Bigg[1- \dfrac{3+\re^2}{1+\re^2}f_r \Bigg] 
\label{eq:D1appen} \\
\Delta_E^{(2)} & = & \dfrac{3+\re^2}{1+\re^2} \dfrac{\Delta^2}{4Un}
\Bigg[\frac{f_r^2}{2} + \left(f_r-\frac{3}{2}f_r^2\right)f_{2r}\Bigg] -\dfrac{2\re^2}{(1+\re^2)^2}\dfrac{\Delta^2 f_r^2}{\re E+2\J}
\label{eq:D2appen} \\
\Delta_E^{(0)} & = & \dfrac{3+\re^2}{1+\re^2} \dfrac{\Delta^2}{4Un}( f_r-f_r^2)-\dfrac{2\re^2}{(1+\re^2)^2}\dfrac{\Delta^2 f_r^2}{\re E+2\J}.
\label{eq:D0appen}
\end{eqnarray}

\end{document}